\documentclass[prl,twocolumn]{revtex4}
\usepackage{graphicx}
\newcommand{\be}{\begin{equation}}
\newcommand{\tc}{{T_{\rm c}}}
\newcommand{\phdag}{{\phantom{\dagger}}}
\newcommand{\ee}{\end{equation}}
\newcommand{\bea}{\begin{eqnarray}}
\newcommand{\eea}{\end{eqnarray}}
\newcommand{\bse}{\begin{subequations}}
\newcommand{\ese}{\end{subequations}}
\newcommand{\kf}{k_{\rm F}}
\newcommand{\bk}{{\bf k}}
\newcommand{\nf}{n_{\rm F}}
\newcommand{\as}{a_{\rm s}}

\newcommand{\curO}{{\cal O}}

\begin{document}

\def\K{{\bf{K}}}
\def\Q{{\bf{Q}}}
\def\Gbar{\bar{G}}
\def\tk{\tilde{\bf{k}}}
\def\k{{\bf{k}}}
\def\q{{\bf{q}}}

\title{Dynamical Cluster Quantum Monte Carlo Study of the Single Particle Spectra of 
Strongly Interacting Fermion Gases}

\author{Shi-Quan Su, Daniel E. Sheehy, Juana Moreno, and Mark Jarrell}
\affiliation{ Department of Physics and Astronomy, Louisiana State
University, Baton Rouge, Louisiana 70803 }

\date{December 16, 2009}

\begin{abstract}

We study the single-particle spectral function of
resonantly-interacting fermions in the unitary regime, as described
by the three-dimensional attractive Hubbard model in the dilute limit.
Our approach, based on the Dynamical Cluster Approximation and the
Maximum Entropy Method, shows the emergence of a gap with decreasing
temperature, as reported in recent cold-atom photoemission
experiments, for coupling values that span the BEC-BCS crossover.
By comparing the behavior of the spectral function to that of the
imaginary time dynamical pairing susceptibility, we attribute the
development of the gap to the formation of local bound atom pairs.

\end{abstract}

\maketitle

%==========BODY OF PAPER =========================================

%{\em Introduction} 
%
One of the most exciting recent developments in correlated systems has
been the observation of pairing and superfluidity of
ultracold atomic fermions  interacting via an s-wave Feshbach resonance~\cite{Gurarie,Giorgini08,Ketterle}.  
The precise tunability of the Feshbach resonance mediated
interaction allows the exploration of superfluidity for coupling extending
from the BCS regime, with loosely-bound Cooper pairs, 
to the BEC 
regime, with local pairs at strong attraction.
Thus such experiments provide a highly controllable setting
to study phenomena that appears in a range of other systems, including
neutron stars~\cite{Gezerlis08}, quark matter~\cite{Alford},
and correlated
electronic systems such as the high temperature
superconductors~\cite{Chen2005}.

While the low-temperature properties of ultracold fermion gases are relatively well understood
and essentially described by the BEC-BCS crossover wavefunction~\cite{Giorgini08}, the
finite  temperature behavior still presents several
mysteries, particularly in the  unitary regime, where the
s-wave scattering length $\as$ between the two
species of fermion diverges.
 At strong couplings
preformed pairs are predicted to exist above the
superfluid transition temperature~\cite{Chen2005},   
possibly yielding a {\it pseudogap} in the density of states~\cite{Randeria}.
But, how the pseudogap phenomena  is reflected in  experimental probes, such as 
radio frequency spectroscopy~\cite{Chin2004,Schunck2008}, is still an open question.

To shed light on the spectral properties of resonantly-interacting fermion gases,
we compute the single particle spectral function $A(\bk,\omega)$ 
as a function of frequency $\omega$ and wavevector $\bk$.
The spectral function has recently been measured in experiments with 
trapped $^{40}$K atoms via a novel momentum-resolved generalization~\cite{Stewart08} of
radio frequency spectroscopy~\cite{Chin2004,Schunck2008}. 
Various analytical methods have been developed to compute  $A(\bk,\omega)$
in the BEC-BCS crossover based on the ladder approximation~\cite{Massignan,Tsuchiya},   
self-consistent many-body theories~\cite{Haussmann07,Chen09,Haussmann09} 
and large-$N$ expansions~\cite{Veillette08}. These approaches are hampered by
the lack of a small parameter in the unitary regime. 

In this letter we combine the Dynamical Cluster Approximation~\cite{hettler:dca,maier:rev} 
and the Maximum Entropy Method~\cite{jarrell:mem} 
to obtain the spectral function in the thermodynamic limit without uncontrollable approximations 
(see also Ref.~\cite{Bulgac08}).
We observe a strong suppression of low-energy spectral weight in conjunction with the
expected onset of pairing correlations. Our simulations also capture 
the concomitant formation of a double-peak structure  in $A(\bk,\omega)$
with decreasing temperature in the strong-coupling regime.  At unitarity, we find
a momentum distribution $n_k$ consistent with the predictions of Tan~\cite{Tan}, and an estimate 
of the universal
parameters $\xi$ and $\zeta$ characterizing the unitary Fermi gas~\cite{BulgacBertsch}.

\begin{figure*}[t]
\begin{center}
\includegraphics*[width=18cm]{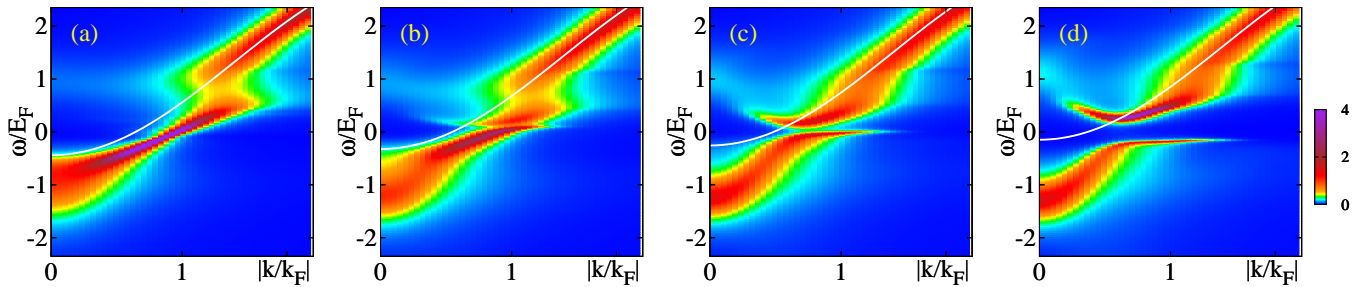}
\caption{(color online) The spectral function
$A(k,\omega)$ of the three-dimensional attractive Hubbard model with filling $\langle
n\rangle=0.3$, at temperature $T/E_F=0.049$, and several coupling strengths.   
(a) On the BCS side of the Feshbach resonance, $U=-6.88t$, $1/(\as \kf)=-0.99$. (b) At
unitary coupling, $U=U_c=-7.92t$, $1/(\as \kf)\longrightarrow 0$. (c)
On the BEC side of the Feshbach resonance, $U=-8.4t$, $1/(\as
\kf)=0.38$. (d) In the BEC region, $U=-9.2t$, $1/(\as \kf)=0.92$. In each
panel, the white curve is the
noninteracting dispersion $\epsilon_{\mathbf{k}}-\mu$.
% and the black curve includes the Hartree term: $\epsilon_{\mathbf{k}}-\mu+ U \langle n \rangle/2$. 
} 
\label{fig:Akw}
\end{center}
\vskip-0.6cm
\end{figure*}

The Dynamical Cluster Approximation (DCA), unlike conventional QMC, 
% usually studying the finite size lattice system) that 
allows the extraction of the single-particle spectral function $A(\k,\omega)$ in the thermodynamic limit.
One disadvantage of the DCA in the present setting
is that it applies to lattice models, while cold atoms possess the itinerant-fermion
dispersion $\epsilon_\bk = k^2/2m$.  However, we address this by focusing
on the {\it dilute limit\/} of the three-dimensional cubic Hubbard model
\begin{eqnarray}
\hat{H} =-t\sum_{\langle i,j \rangle,\sigma}
\hat{c}^\dagger_{i\sigma}\hat{c}^\phdag_{j\sigma}+U\sum_i\hat{n}_{i\uparrow}\hat{n}_{i\downarrow}
-\mu\sum_{i\sigma}\hat{n}_{i\sigma},
\label{eq:Hub}
\end{eqnarray}
where $\hat{c}_{i\sigma}$ annihilates a spin-$\sigma$ fermion at site $i$, 
the angle brackets restrict the sum over nearest neighbor
sites with the matrix element $t$, and $U$ is the attractive
on-site interaction.  Here, the chemical potential $\mu$ is tuned to
keep the filling at the small value $\langle n\rangle =0.3$, where
$\langle n\rangle =1$ is half filling. The Fermi energy associated with this filling 
is $E_F=3.414t$.
In this dilute limit, the
hopping dispersion $\epsilon_{\mathbf{k}}=2t[3-\sum_{i=x,y,z}\cos k_i]$ can be approximated 
by its low $\bk$ expansion $\epsilon_k \simeq tk^2$, thus describing free fermions if we take $t = 1/2m$.

By tuning $U$ close to $U_c = -7.92t$, the interaction strength where a two-body
bound state forms in vacuum~\cite{Burovski06}, 
we can capture the unitary regime in cold Fermi gases.  
We expect that interaction effects are insensitive to the underlying lattice when the
scattering length
\be
\as = \frac{UU_c}{8\pi t (|U|-|U_c|)},
\ee
exceeds the lattice spacing.
We find, as discussed below, 
that Eq.~(\ref{eq:Hub}) works well in the BCS and unitary regime, but begins
to deviate from the free-fermion results in the deep-BEC regime of tightly-bound
{\it local\/} pairs where lattice effects cannot be neglected. 
 
%{\em{Formalism.}} 
The DCA covers the lattice with clusters of $N_c=L_c^3$ sites
using periodic boundary conditions. 
Correlations up to length $L_c$ are considered exactly while 
long range correlations are treated at the mean-field level~\cite{hettler:dca,maier:rev}.
We use the Hirsch-Fye quantum Monte Carlo (QMC) algorithm~\cite{hirsch86} as the quantum
solver~\cite{jarrell:dca}, and the maximum entropy method~\cite{jarrell:mem} 
to analytically continue the imaginary-time spectra to real frequencies. In the DCA the momentum
resolution of the self energy $\Sigma$ is determined by a set of $N_c$
$K$-points corresponding to the cluster reciprocal space\cite{maier:rev}. We use a Fourier
 Transform interpolation scheme to approximate $\Sigma$ on other k-points in
the Brillouin zone. We use a Betts cluster\cite{Betts99} with $N_c=24$. 
Our Hirsch-Fye QMC algorithm is efficient as long
as $|U|$ is below the bandwidth $12t$, allowing us
to study the BEC-BCS crossover but not the asymptotic BEC limit.

%{\em{Results.}} 
 We have obtained the single-particle spectral function 
$\displaystyle A(\bk,\omega) = - \frac{1}{\pi} {\rm Im} G(\bk,\omega)$, with $G(\bk,\omega)$ the
retarded Green's function,
across the BEC-BCS crossover and for a wide range of temperatures.
Our displayed results are along the line connecting the center of the
first Brillouin zone  ${\bf\Gamma}=(0,0,0)$ to the corner
${\bf K}=(\pi,\pi,\pi)$, with $k$ referring to the wavevector magnitude. 
The spectral function along other directions is qualitatively equivalent.
We focus on the large scattering length limit with $\displaystyle -1\alt \frac{1}{\kf \as} \alt 1$, where 
we approximate the Fermi wavelength $\kf$ by the itinerant-fermion formula $\kf = (3\pi^2n)^{1/3}$.  
Our simulations are in the normal state, meaning we cannot fully describe the broken symmetry state 
below the superfluid transition at $\tc$.  Above $\tc$, however, 
expect our results to be quantitatively accurate.
Previous work~\cite{Burovski06} has demonstrated that Eq.~(\ref{eq:Hub}) yields reliable results
for the transition temperature of itinerant superfluid Fermi gases. By taking the extrapolated
transition temperature $\tc$  from Ref.~\onlinecite{Burovski06} we get $\tc \simeq 0.064 E_F$
at unitarity coupling and filling $\langle n \rangle = 0.3$.  

The most striking feature of our results,
shown in Fig.~\ref{fig:Akw}, is the strong suppression of the spectral
weight near zero frequency 
with increasing attraction towards the BEC limit.
Fig.~\ref{fig:Akw} displays results for four coupling values at the 
temperature $T/E_F = 0.049$.  In all plots, the behavior at high-$k$ is
an artifact of the underlying lattice that we shall ignore; this is reasonable 
since experiments only probe the occupied part of the spectral function. 
Fig.~\ref{fig:Akw}a, in the weak-coupling BCS regime
$\displaystyle \Big( \frac{1}{\kf \as} = -.99\Big)$, shows an essentially free-particle dispersion for 
$k<\kf$ and $\omega<0$.    The narrowing
of the linewidth as the Fermi surface is approached from $k\to \kf^-$ is clearly seen, consistent
with the system being a Fermi liquid in this regime.  
With increasing coupling a two branch structure 
 begins to form at the Fermi surface.  While this behavior is barely visible at the unitary
point, $\displaystyle \frac{1}{\kf \as} =0$, shown in Fig.~\ref{fig:Akw} (b),
it increases in magnitude with increasing interaction strength, as seen in 
panels (c) and (d).  In Fig.~\ref{fig:Akw}(d), the deep BEC limit,
the two branches of the spectral function are clearly visible, and qualitatively
consistent with the recent cold-atom photoemission experiments~\cite{Stewart08}.

\begin{figure}[t]
\begin{center}
\includegraphics*[width=8cm]{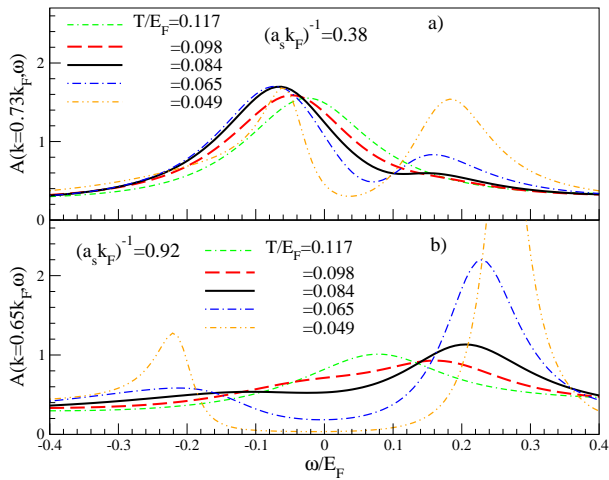}
%\begin{figure} [bth]
%\includegraphics[width=9.0cm,height=6.0cm]{dos.eps}
\caption{(color online) Plots of $A(k,\omega)$ at fixed
momenta %$\bk\alt \kf$ 
as a function of frequency at successive 
decreasing temperatures, for  a) Close to unitarity  
$\displaystyle \Big(  \frac{1}{\kf \as}=0.38 \Big) $ and b) in the deep BEC regime 
$\displaystyle\Big(  \frac{1}{\kf \as}=0.92 \Big) $.  Since we expect superfluidity only below 
$\tc \alt 0.064E_F$~\cite{Burovski06}, the suppression of spectral weight above this temperature
represents the pseudogap. 
 }
\label{fig:edc}
\end{center}\vskip-0.6cm
\end{figure}

\begin{figure}[t]
\begin{center}
\includegraphics*[width=8cm]{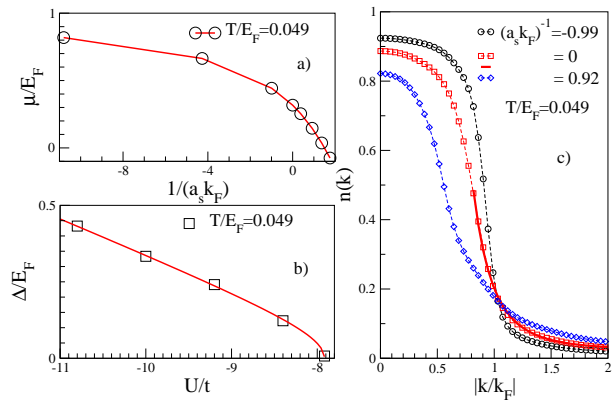}
\caption{(color online) (a) Chemical potential $\mu$ as a function of coupling 
at $T/E_F=0.049$.
 (b) The pseudogap $\Delta$ extracted from
$A(\bf{k},\omega)$ (square symbols). 
%The solid line is the fitting function
%$\displaystyle \Delta/t=\frac{1}{2}\sqrt{1.25(|U|-|U_c|)/t+0.61(|U|-|U_c|)^2/t^2} $. 
(c)
Momentum distribution $n(\bk)$ for different values of $1/(\as \kf)$.
The dashed lines are used to guide the eye, and the unitarity curve
has been fit at large $\bk$ to the Tan relation~\cite{Tan}.}
\label{fig:mugapnk}
\end{center}\vskip-0.6cm
\end{figure}

To examine the onset of the pseudogap with decreasing temperature, we
plot $A(k,\omega)$ at fixed $k$ as a function of energy in Fig.~\ref{fig:edc}.
For each coupling we have chosen the value of momentum $k$ for which the distance between the 
two spectral branches is the shortest.
We find the pseudogap, with decreasing $T$,  only for coupling values beyond the unitary point.
In the BEC regime, as seen in Fig.~\ref{fig:edc}, the high temperature 
spectral functions show a Gaussian shape at high temperatures that develops a pseudogap 
at lower temperatures. The pseudogap separates two 
asymmetric peaks, reminiscent of recent analytic results~\cite{Tsuchiya}. 
We define the pseudogap $\Delta$ as one-half the minimum peak-to-peak distance 
between the two spectral function peaks; our values for $\Delta$ at the lowest 
temperature (characterizing the evolution of the pseudogap across the BEC-BCS crossover)
are plotted in Fig.~\ref{fig:mugapnk}(b).
%
% we provide a prediction for the shape of the 
%pseudogap across the BEC-BCS crossover. 
%Our values for $\Delta$, at the lowest temperatures, are 
%plotted in Fig.~\ref{fig:mugapnk}(b) for the coupling values where we see a pseudogap form
%with decreasing temperature. 

Even though our simulation  does not explore the superfluid state, it is able to 
capture the onset of pairing correlations and estimate the universal
parameters $\xi$ and $\zeta$ characterizing the unitary Fermi gas~\cite{BulgacBertsch}.
For example, the internal energy per particle can be expressed as
\be
E/N = \frac{3}{5}E_F \big[\xi  - \frac{\zeta}{\kf \as} + \curO\big(\frac{1}{(\kf\as)^2}\big)\big].
\label{formula}
\ee
near unitarity~\cite{BulgacBertsch,Carlson2003}.
We quantify $\xi$ from our extracted values of $\mu$, that is adjusted to always keep filling
$\langle n \rangle = 0.3$. Fig.~\ref{fig:mugapnk}(a) displays $\mu$  as a function of coupling.
The chemical potential is positive and close to $E_F$ in the BCS regime and reduces monotonically 
with increasing attraction towards the  BEC limit of local pairing.  We find $\mu/E_F = 0.31$ at
unitarity, yielding our prediction for the universal parameter  $\mu/E_F = \xi=0.31$, which is 
consistent with recent experimental and theoretical results, 
see e.g. Refs.~\cite{Giorgini08,Veillette07}.

We  also calculate 
the momentum distribution
\begin{eqnarray}
n({k}) = \int^{\infty}_{-\infty} d \omega \nf(\omega)A(k,\omega),
\end{eqnarray}
with $\displaystyle \nf(\omega) = \frac{1}{1+{\rm e}^{\beta\omega}}$ the Fermi function.
The momentum distribution  has been previously measured
in the BEC-BCS crossover~\cite{RegalMOM}.  
Fig.~\ref{fig:mugapnk}(c) displays our results for $n({k})$ showing 
the expected behavior: a sharp step in $n(k)$ 
in the BCS regime that broadens with increasing attraction towards the BEC limit.  
To gain additional insight, we fit these curves to $c_0+ c_1/(|k/\kf|+c_2)^4$ to test whether the 
Tan relation~\cite{Tan} 
\be
n(k) =C/k^4, \,\,\, {\rm for\/}\,\,\, k\to \infty,
\label{eq:tan}
\ee
governing the large $k$ behavior of the momentum occupation, holds for our system.  
Although our system mimics a free-fermion dispersion at small momenta, at large $k$ lattice effects 
become inescapable,
and it is not clear if the Tan relation will be valid since it probes the 
extremely short distance behavior of interacting fermions.   Despite this, we find that,
at unitarity, the optimal fitting parameters are $c_0 =0.019$, $c_1 = 0.148$ 
and $c_2 = -0.057$. The  $c_0$ and $c_2$  are reasonably small and
the fit, shown as a solid curve for the unitary plot in Fig.~\ref{fig:mugapnk}(c), is seen to be quite accurate, 
showing that our system does exhibit the universal properties of a unitary Fermi gas.
From the theoretical relation $c_1 = 2\zeta/5\pi$,
with $\zeta$ governing the interaction dependence of the energy per particle 
near unitary~\cite{Tan} (Eq.~(\ref{formula})), we extract the prediction $\zeta \simeq 1.16$. 
In the BCS and BEC regimes, the best-fit values of $c_0$ and $c_2$ are larger, so that we do not find a 
good fit to the Tan formula. 
In particular in the deep BEC region, one expects
significant  occupation of high momentum states.  While this certainly occurs in our system, the finite
size of our momentum-space Brillouin zone restricts the possible occupied momenta, preventing our
system from truly accessing the deep BEC limit.

It is natural to assume that the pseudogap observed in 
our single-particle spectra arises from the formation of s-wave pairing 
correlations. %in a strongly interacting Fermi gases.  
To test this assumption, we study the local s-wave pair susceptibility 
$ \chi^s(R,\tau)$ at spatial distance $R=i-i'$ and imaginary time $\tau$.  Setting $R=0$ and
$\tau = \beta/2$, we have 
\be
\chi^s(0,\beta/2) =  \frac{1}{N_c}\sum_i\langle \Delta^\dagger_s(i,\beta/2)\Delta^\phdag_s (i',0)\delta_{i,i'}+\it{h.c.} \rangle,
\ee
where $\Delta^s(i,\tau)=c_{i\uparrow}(\tau)c_{i\downarrow}(\tau)$  is the 
s-wave pair operator and the average occurs over system sites. This quantity, $\chi^s(0,\beta/2)$,  
probes local pairing correlations with the longest distance 
along the imaginary time direction. 

\begin{figure}[t]
\begin{center}
\includegraphics*[width=8cm]{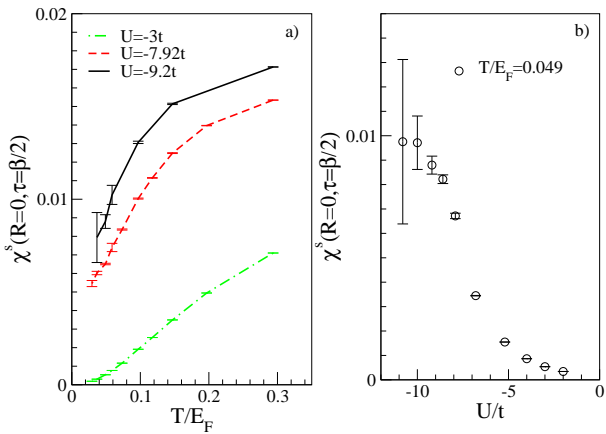}
\caption{(color online) Local on-site pairing susceptibility
$\chi^s(R=0,\beta/2)$ (a) as function of temperature $T$ for different couplings   and (b) as
function of coupling $U$ for $T/E_F=0.049$. } \label{fig:chiTu}
\end{center}
\vskip-.6cm
\end{figure}

In Fig.~\ref{fig:chiTu}, we plot $\chi^s(0,\beta/2)$ as
function of temperature (left panel) and coupling (right
panel). On panel (a), we can see that $\chi^s(0,\beta/2)$
vanishes when $T\rightarrow 0$ at weak coupling. This
signals that, in the weak coupling region, atoms can not form stable local pairs; although, of course, 
Cooper pairing correlations will set in at very much lower temperatures, outside the range of our 
computational results. When $|U|$ is large enough, towards the BEC limit, $\chi^s(0,\beta/2)$
shows a {\em finite\/} $U$-dependent intercept in the zero temperature limit,  showing
the establishment of local pairs. Fig. ~\ref{fig:chiTu}(b) shows
the coupling dependence of $\chi^s(0,\beta/2)$ at low temperatures.  While this 
quantity is small at weak coupling in the BCS regime, near unitarity it 
undergoes a rapid increase as local pairing is established.

%{\em{Discussion and Conclusions.}} 
To summarize, we have demonstrated that the Dynamical Cluster Quantum Monte Carlo, along with the
Maximum Entropy Method, 
may be used to make quantitatively-reliable predictions for the single-particle spectral
function of unitary Fermi gases.  Our spectral function results capture features seen in
recent  $^{40}K$ photoemission experiments~\cite{Stewart08}.  We find the onset of a pairing pseudogap 
with decreasing temperature (above the expected superfluid transition temperature) for coupling
values beyond the unitary point in the BEC regime.
 Since our 
method treats the short range correlations of the system exactly, it
holds particular promise for obtaining quantitatively accurate predictions for radio frequency spectra
and photoemission data in the strongly-interacting unitary regime of superfluid Fermi gases. 
Future work will extend this method to incorporate the effect of a background trapping 
potential, a necessary step to quantitatively understand experimental results.

\acknowledgments We acknowledge useful discussions with Dana Browne, Ehsan Khatami, Karlis Mikelsons, Unjong Yu, and Zhaoxin Xu. This research was supported by an allocation of computing time from the Ohio Supercomputer Center, by the NSF OISE-0952300 and the Louisiana Board of Regents, under grant No. LEQSF (2008-11)-RD-A-10.

\end{document}